\documentclass[12pt]{article}
\usepackage{amssymb,amsmath}
\usepackage[noblocks]{authblk}
\usepackage[top=0.75in, bottom=0.75in, left=0.75in, right=0.75in, dvips]{geometry}
\usepackage{caption}
\usepackage{graphicx}
\usepackage{epstopdf}
\begin{document}
\textwidth 10.0in
\textheight 9.0in
\topmargin -0.60in\title{A Systematic Expansion of Running Couplings and Masses}
\author[1,*]{F.A. Chishtie}
\author[1,2]{D.G.C. McKeon}
\author[3]{T.N. Sherry}
\affil[1] {Department of Physics and Astronomy, The
University of Western Ontario, London, ON N6A 5B7, Canada}
\affil[2] {Department of Mathematics and
Computer Science, Algoma University,\newline Sault Ste. Marie, ON P6A
2G4, Canada}
\affil[3] {School of Mathematics, Statistics and Applied Mathematics, National University of Ireland Galway, University Road, Galway H91 TK33 Ireland} 
\affil[*]{Corresponding author}
\maketitle                              
   
\maketitle
\noindent
PACS No.: 11.10Hi\\
Key Words: Running parameters, Renormalization group summation\\
email: fchishti@uwo.ca, dgmckeo2@uwo.ca, tom.sherry@nuigalway.ie

\begin{abstract}
As an alternative to directly integrating their defining equations to find the running coupling $a(\mu)$ and the running mass $m(\mu)$, we expand these quantities in powers of $\ln\left(\frac{\mu}{\mu^\prime}\right)$ and their boundary values $a(\mu^\prime)$ and 
$m(\mu^\prime)$.  Renormalization group summation is used to partially sum these logarithms.  We consider this approach using both the $\overline{MS}$ and 't Hooft renormalization schemes. We also show how the couplings and masses in any two mass independent renormalization schemes are related.
\end{abstract}

Two essential ingredients of quantum chromodynamics are the running coupling $a(\mu)$ and the running mass $m(\mu)$, which when using mass independent renormalization [1,2] satisfy
\begin{subequations}
\begin{align}
\mu \frac{da(\mu)}{d\mu} &= \beta(a) = -ba^2\left( 1 + ca + c_2a^2 + c_3a^3 + \ldots\right)\\
\mu \frac{dm(\mu)}{d\mu} &= m\gamma(a) = m fa\left( 1 + g_1a + g_2a^2 + \ldots \right),
\end{align}
\end{subequations}
where $c_i$, $g_i$ requires an $(i + 1)$ loop calculation [3].  Under a change of renormalization scheme (RS) corresponding to a finite renormalization of the coupling and mass
\begin{subequations}
\begin{align}
\overline{a} &= a + x_2a^2 + x_3 a^3 + \ldots \\
\overline{m} &= \left( 1 + y_1 a + y_2 a^2 + \ldots\right)
\end{align}
\end{subequations}
only the coefficients $b$, $c$, $f$ are unaltered [4].  Using $\overline{MS}$, the coefficients $g_i\left(i = 1,2,3,4\right)$, $c_i(i = 2,3,4)$ have been evaluated [11,12].  In this paper we show how one can systematically obtain useful perturbative expressions for $a(\mu)$ and $m(\mu)$.  The approach used also appears in refs. [4].  This provides an alternative to explicitly integrating the formal solutions to eq. (1,2)
\begin{subequations}
\begin{align}
\ln \left(\frac{\mu}{\Lambda}\right) &= \int_0^{a(\mu)} \frac{dx}{\beta(x)} + 
\int_0^\infty \frac{dx}{bx^2(1+cx)}\\
m(\mu) &= I\!\!M \exp \left( \int_0^{a(\mu)} \frac{dx \gamma(x)}{\beta(x)} + 
\int_0^\infty \frac{dxfx}{bx^2(1+cx)} \right)
\end{align}
\end{subequations}
where $\Lambda$ and $I\!\!M$ are related to the boundary values of $a$ and $m$, consistent with the conventions of ref. [5,6].

In our approach, we utilize the expansions
\begin{subequations}
\begin{align}
a^\prime  &= a\left[ 1 + \left(\alpha_{11} \ell\right) a + \left(\alpha_{21} \ell + \alpha_{22}\ell^2\right) a^2 +  \ldots\right] \\
m^\prime  &= m\left[ 1 + \left(\beta_{11} \ell\right) a + \left(\beta_{21} \ell + \beta_{22}\ell^2\right) a^2 +  \ldots\right]
\end{align}
\end{subequations}
where $a^\prime = a(\mu^\prime)$, $m^\prime = m(\mu^\prime)$; $a = a(\mu)$, $m = m(\mu)$ and $\ell = \ln\left(\frac{\mu}{\mu^\prime}\right)$. One should think of $a^\prime$ and $m^\prime$ in eq. (4) as solutions to eq. (1), expanded in terms of boundary values of $a$ and $m$. Since $a^\prime$ and $m^\prime$ are independent of $\mu$, we can say that
\begin{subequations}
\begin{align}
\mu \frac{da^\prime}{d\mu} &= 0 =  \left(\mu \frac{\partial}{\partial\mu} + 
\beta(a) \frac{\partial}{\partial a}\right)a^\prime \\
\mu \frac{dm^\prime}{d\mu} &= 0 =  \left(\mu \frac{\partial}{\partial \mu} + 
\beta(a) \frac{\partial}{\partial a} + m\gamma(a) \frac{\partial}
{\partial m}\right)m^\prime .
\end{align}
\end{subequations}
If, in eq. (4), we define
\begin{subequations}
\begin{align}
a^\prime & = \sum_{n=0}^\infty S_{n} (al)  a^{n+1}\\
m^\prime & = m\sum_{n=0}^\infty  T_{n} (al)  a^{n}
\end{align}
\end{subequations}
where $S_n(\xi)= \sum_{k=0}^\infty \alpha_{n+k,k} \xi^k$, $T_n(\xi) = \sum_{k=0}^\infty \beta_{n+k,k} \xi^k$ and $\alpha_{n,0}=\beta_{n,0}=\delta_{n,0}$, we find that eqns. (5a,b) take the form
\begin{subequations}
\begin{align}
\left(  \mu \frac{\partial}{\partial\mu} + 
\beta(a) \frac{\partial}{\partial a}\right)\sum_{n=0}^\infty S_n(a\ell)a^{n+1} &= 0\\
\left(  \mu \frac{\partial}{\partial\mu} + 
\beta(a) \frac{\partial}{\partial a}+ m \gamma (a) \frac{\partial}{\partial m} \right) \sum_{n=0}^\infty  m T_n(a\ell)a^n &= 0.
\end{align}
\end{subequations}
Together, eqs. (1,7) result in a pair of sets of nested equations - one set for $S_n$, the first two of which are
\begin{subequations}
\begin{align}
(1-b\xi)S_0^\prime &- b S_0 = 0\\
(1-b\xi)S_1^\prime &- 2bS_1 - bc \left( S_0^\prime \xi + S_0 \right) = 0;
\end{align}
\end{subequations}
and the second set for $T_n$, the first two of which are
\begin{subequations}
\begin{align}
(1-b\xi) T_0^\prime + fT_0 & = 0\\
(1-b\xi)T_1^\prime  + (f-b)T_1 & - bc  T_0^\prime \xi + fg_1 T_0  = 0 .
\end{align}
\end{subequations}
In general the equation for $S_n$ requires knowing $(S_0 \ldots S_{n-1})$ as well as $(b,c,c_2 \ldots c_n)$ and for $T_n$ requires knowing $(T_0 \ldots T_{n-1})$ as well as $(b,c,c_2 \ldots c_n, f, g_0 \ldots g_n)$. Using the boundary conditions $S_n(0) = T_n(0) = \delta_{n,0}$ we illustrate the solutions of eqs. (8,9) as follows: 
\begin{subequations}
\begin{align}
S_0 &= \frac{1}{w} \qquad (w = 1 - b\xi)
\end{align}
\begin{align}
S_1 &= \frac{-c \ln w}{w^2}
\end{align}
\begin{align}
S_2 &= \frac{c^2\left(\ln^2 w - \ln w + w-1\right)-c_2(w-1)}{w^3}
\end{align}
\begin{align}
S_3 &= \frac{1}{2w^4}\Big[ -2c^3 \ln^3w + 5c^3\ln^2w - 4 \left\{(-1+w)c^2 - c_2 \left(w - \frac{3}{2}\right)\right\}c\ln w\nonumber \\
&\hspace{2cm} - (-1+w) \left\{(-1+w)c^3 -2cc_2w + c_3(w+1)\right\}\Big]
\end{align}
\begin{align}
S_{{4}}=\frac{1}{6{w}^{5}}\Big[6\,{c}^{4} \ln ^{4} w -26\,{c}^{4} \ln ^{3} w
+ \left\{ 18\,{c}^{2}\left( {c}^{2}-c_{{2}} \right) w-9\,{c}^{2} \left( {c}^{2}-4\,c_{{2}}\right)  \right\}   \ln^{2}w \nonumber \\
+ \left\{ 6\,c \left( {c}^{3}-2\,cc_{{2}}+c_{{3}} \right) {w}^{2}-30\,{c}^{2}\left( {c}^{2}-c_{{2}} \right) w+6\,c \left( 4\,{c}^{3}-3\,cc_{{2}}-2\,c_{{3}} \right)  \right\} \ln w\nonumber \\
+ \left( 2\,{c}^{4}-6\,c_{{2}}{c}^{2}+4\,c_{{3}}c+2\,{c_{{2}}}^{2}-2\,c_{{4}} \right) {w}^{3}+ \left( 3\,{c}^{4}-6\,c_{{2}}{c}^{2}-3\,c_{{3}}c+6\,{c_{{2}}}^{2}\right) {w}^{2}\nonumber \\
- \left( 12\,{c}^{2}-18\,c_{{2}} \right)  \left( {c}^{2}-c_{{2}} \right) w+7\,{c}^{4}-18\,c_{{2}}{c}^{2}-c_{{3}}c+10\,{c_{{2}}}^{2}+2\,c_{{4}}\Big]
\end{align}
\begin{align}
T_0 &=  w^\rho \qquad ( \rho = f/b) 
\end{align}
\begin{align}
T_1 &= -\rho w^{-1+\rho} \left[ -c \ln w + (c-g_1) (w-1)\right]
\end{align}
\begin{align}
T_2 &= \frac{\rho}{2} w^{-2+\rho} \bigg[ -c^2 (1-\rho) \ln^2 w + \left\{ -2  \rho (c-g_1)(w-1) + 2g_1\right\} c \ln w\\
&+ (w-1) \left\{ \left( c^2 - c_2 + \rho (c-g_1)^2\right) (w-1) + \left( g_2 - cg_1 \right) (w+1) \right\} \bigg]. \nonumber
\end{align}
\end{subequations}

For $T_3$ and $T_4$ see Appendix A. \\

It is possible to verify that if eq. (10) is used to express $a(\mu^{\prime\prime})$ in terms of $a(\mu^\prime)$ and then $a(\mu^\prime)$ is expressed in terms of $a(\mu)$, one obtains what is expected for $a(\mu^{\prime\prime})$ in terms of $a(\mu)$. Furthermore, it is also possible to show that
\begin{subequations}
\begin{align}
\mu^\prime \frac{d}{d\mu^\prime} a(\mu^\prime) &= \beta(a(\mu^\prime))\\
\mu^\prime \frac{dm(\mu^\prime)}{d\mu^\prime}  &= m(\mu^\prime)\gamma(a(\mu^\prime)).
\end{align}
\end{subequations}
These two consistency checks are most easily verified if the expansion coefficients $\alpha_{mn}$, $\beta_{mn}$ of eq. (4) are used.

When using the 't Hooft renormalization scheme [17] we set $c_i = 0 (i \geq 2)$ and $g_i =0 (i \geq 1)$.  In this case the coupled equations for $S_n$ and $T_n$ simplify, (where $ \dot{f} \equiv \frac{d}{dw}  f(w)$
\begin{subequations}
\begin{align}
\dot{S}_n +  \frac{(n+1)}{w} S_n + c \left[ \left( 1 - \frac{1}{w}\right) \dot{S}_{n-1} + \frac{n}{w} S_{n-1} \right] & = 0\\
\dot{T}_n +  \frac{(n - \rho)}{w} T_n + c \left[ \left( 1 - \frac{1}{w}\right) \dot{T}_{n-1} + \frac{n-1}{w} T_{n-1} \right] & = 0.
\end{align}
\end{subequations}

Upon setting $w^{n+1}S_n = c^n\sigma_n$, eq. (11a) leads to 
\begin{subequations}
\begin{align}
\sigma_n (w) & = f_{n-1}(w) - (-1)^{n+1} - n \int_1^w \frac{dx}{x} f_{n-1}(x) \quad (n = 1,2 \ldots)\\
\intertext{\rm{where} $\sigma_0 = 1$ \rm{and}}
f_{n-1}(x) & = \sigma_{n-1} (x) - x\sigma_{n-2} (x) + x^2 \sigma_{n-3}(x) + \ldots + (-1)^{n+1} x^{n-1} \sigma_0(x)
\end{align}
\end{subequations}
while from eq. (11b), if $w^{n-\rho}T_n = c^n\tau_n$, it follows that $(n = 0,1, \ldots)$
\begin{equation}
\frac{d}{dw} \tau_n + (w-1) \frac{d}{dw} \tau_{n-1} + \rho \left(1 - \frac{1}{w}\right) \tau_{n-1} + \frac{n-1}{w}\tau_{n-1} = 0.
\end{equation}
It thus proves relatively easy to obtain $S_n$ and $T_n$ in the 't Hooft renormalization scheme. In particular, the solutions for $T_3$ and $T_4$ in the 't Hooft RS are significantly simpler than the solutions in a general RS provided in Appendix A:
\begin{subequations}
\begin{align}
T_{{3}}= \frac{1}{6}{w}^{\rho-3}\rho\,{c}^{3}( \ln w  -w+1 )\Big[ ( \rho-1 )  ( \rho-2 )  \ln^{2}w + ( - ( 2\,\rho-2)  ( \rho+1 ) w+2\,{\rho}^{2}-5 ) \ln w   \nonumber \\
+ ( \rho+2 )  ( \rho+1 ) {w}^{2}+ ( -2\,{\rho}^{2}-6\,\rho-1) w+{\rho}^{2}+3\,\rho-1\Big]
\end{align}

\begin{align}
T_{{4}}= \frac{1}{24}\,{w}^{\rho-4}\rho\,{c}^{4}( \ln w  - w+1 )\Big[( \rho-1 )  ( \rho-2 ) ( \rho-3) ( \ln ^{3} w +( - ( 3\,\rho-3 )  ( \rho-2 )  ( \rho+1) w \nonumber \\ 
+3\,{\rho}^{3}-6\,{\rho}^{2}-15\,\rho+26)  (\ln^{2}w +\{ ( 3\,\rho-3 ) ( \rho+2 )  ( \rho+1 ) {w}^{2}+ ( -6\,{\rho}^{3}-12\,{\rho}^{2}+30\,\rho+8 ) w+3\,{\rho}^{3} \nonumber \\ 
+6\,{\rho}^{2}-27\,\rho-2 \} \ln w - ( \rho+3 ) ( \rho+2 )  \left( \rho+1 \right) {w}^{3}+ \left( 3\,{\rho}^{3}+18\,{\rho}^{2}+21\,\rho+2 \right) {w}^{2}\nonumber \\ 
+ \left( -3\,{\rho}^{3}-18\,{\rho}^{2}-9\,\rho+14 \right) w+{\rho}^{3}+6\,{\rho}^{2}-\rho-10)\Big] 
\end{align}
\end{subequations}

In the 't Hooft RS, eqs. (3a,b) can be integrated to obtain
\begin{subequations}
\begin{align}
m &= \frac{I\!\!M f}{b} \ln \left( \frac{1 + ca}{ca}\right)
\intertext{and}
\zeta &+ \ln \zeta = 1 + \frac{b}{c} \ln \left(\frac{\mu}{\Lambda}\right)
\end{align}
\end{subequations}
where $\zeta \equiv 1 + \frac{1}{ca}$ so that the solution for $a\left( \ln \frac{\mu}{\Lambda}\right)$ is in terms of the appropriate Lambert function $w(x) (we^w = x)$.

It is possible to express $\overline{a}$ and $\overline{m}$, the coupling and mass apprpriate to the RS defined by the parameters $\overline{c}_i$ and $\overline{g}_i$, in terms of $a$ and $m$, the coupling and mass appropriate to the RS defined by the parameters $c_i$ and $g_i$.  To do this, we make use of [7,8]
\begin{subequations}
\begin{align}
\frac{\partial a}{\partial c_i} = B_i(a) &= -b\beta(a) \int_0^a dx \frac{x^{i+2}}{\beta^2(x)} \\
&\approx a^{i+1} \left[ \frac{1}{i-1} - c \left( \frac{i-2}{i(i-1)}\right) a + \frac{1}{i+1}\left( c^2 \frac{i-2}{i}- c_2 \frac{i-3}{i-1}\right) a^2 + \ldots \right]\nonumber \\
\frac{\partial a}{\partial g_i} &= 0\\
\frac{1}{m}\frac{\partial m}{\partial c_i} = \Gamma_i^c(a) &=  \frac{\gamma (a)}{\beta(a)} \beta(a) + b \int_0^a dx \frac{x^{i+2}\gamma(x)}{\beta^2(x)} \\
&\approx \rho a^i \bigg[ \frac{-1}{i(i-1)} + 2 \left( \frac{c}{i(i+1)} - \frac{g_1}{(i+1)(i-1)}\right) a \nonumber \\
& + \frac{1}{i+2} \left( \frac{2c_2 - 3c^2}{i+1} + \frac{4g_1c}{i} - \frac{3g_2}{i-1}\right) a^3 + \ldots \bigg]\nonumber \\
\intertext{\rm{and}}
\frac{1}{m} \frac{\partial m}{\partial g_i}&= \Gamma_i^g (a) = f \int_0^a dx \frac{x^{i+1}}{\beta(x)}\\
&= \frac{f}{b} a^i \bigg[ - \frac{1}{i} + \left( \frac{c}{i+1}\right) a + \left( \frac{c_2 - c^2}{i+2} \right) a^2 \nonumber \\
&+ \left( \frac{c_3 + c^3 - 2cc_2}{i+3}\right) a^3 + \left( \frac{c_4-c^4 + 2c^2c_2 - 2cc_3-c_2^2}{i+4}\right) a^4 + \ldots \bigg].\nonumber
\end{align}
\end{subequations}
If we now expand
\begin{subequations}
\begin{align}
\overline{a} &= a\left[1 + \phi_1\left( c_j, \overline{c}_j\right) a + \phi_2 \left(c_j, \overline{c}_j\right) a^2 + \ldots \right]\\
\overline{m} &= m\left[1 + \psi_1\left( c_j, \overline{c}_j;g_j, \overline{g}_j \right) a + \psi_2 \left(c_j, \overline{c}_j; g_j, \dot{g}_j\right) a^2 + \ldots \right]
\end{align}
\end{subequations}
where $\phi_i\left(c_j,c_j; g_j,g_j\right) = 0 = \psi_i\left(c_j,c_j; g_j,g_j\right)$, then the equations
\begin{subequations}
\begin{align}
\frac{d\overline{a}}{dc_i} & = \left( \frac{\partial}{\partial c_i} + B_i (a) \frac{\partial}{\partial a}\right)\overline{a} = 0\\
\frac{d\overline{a}}{dg_i}& = 0\\
\frac{d\overline{m}}{dc_i} & \left( \frac{\partial}{\partial c_i} + B_i(a) \frac{\partial}{\partial a} + m \Gamma_i^c (a) \frac{\partial}{\partial m}\right) \overline{a} = 0\\
\frac{d\overline{m}}{dg_i} & = \left( \frac{\partial}{\partial g_i} + m\Gamma_i^g (a) \frac{\partial}{\partial m}\right)\overline{m} = 0
\end{align}
can be used to write down two sets of nested equations for the coefficients $\phi_i(c_j,\overline{c}_j)$ and $\psi_i(c_j,\overline{c}_j;g_j,\overline{g}_j)$ in (18a) and (18b). 
\end{subequations}
Using the solutions of these nested equations in (18a,b), we find [6]
\begin{subequations}
\begin{align}
\overline{a} &= a\bigg\{ 1 + \left( \overline{c}_2 - c_2\right) a^2 + \frac{1}{2} 
\left( \overline{c}_3 - c_3\right)a^3 \\
 &+ \left[ \frac{1}{3} \left( \overline{c}_4 - c_4\right) - \frac{c}{6}
\left( \overline{c}_3 - c_3\right) + \frac{1}{6} \left( \overline{c}_2^2 - c_2^2\right) + \frac{3}{2} \left( \overline{c}_2 - c_2\right)^2\right] a^4 + \ldots \bigg\}\nonumber \\
\overline{m} & = m \bigg\{ 1 + \rho \left( g_1 - \overline{g}_1\right) a + 
\frac{\rho}{2} \left[ g_2 - \overline{g}_2 + c_2 - \overline{c}_2 - c \left(g_1 - \overline{g}_1\right) + \rho \left(g_1 - \overline{g}_1\right)^2\right]a^2 \nonumber \\
&+ \Big[ -\frac{1}{6} \rho^3 (\overline{g}_1 - g_1)^3 - \frac{1}{2}\rho^2 c (\overline{g}_1 - g_1)^2 + \frac{1}{2}\rho^2 (\overline{c}_2-c_2) + \frac{1}{2} \rho^2 (\overline{g}_2 - g_2)\nonumber \\
&+ \frac{1}{3} \rho (c - 2\overline{g}_1)(\overline{c}_2-c_2) -  \frac{1}{6} \rho (\overline{c}_3 - c_3) -  \frac{1}{3} \rho (c^2 - c_2)(\overline{g}_1 - g_1)\nonumber \\
 &+ \frac{1}{3} \rho c (\overline{g}_2 - g_2) -  \frac{1}{3} \rho (\overline{g}_3 - g_3)\Big]a^3\nonumber \\
&+ \Big[  \frac{1}{8} \rho (\rho + 1)  (\overline{c}_2 - c_2)^2 +
\Big(-\frac{1}{4} \rho^3(\overline{g}_1 - g_1)^2 - \rho(\rho + 1) (\frac{7}{12} c - \frac{2}{3} \overline{g}_1)(\overline{g}_1 - g_1)\nonumber \\
&+ \frac{1}{4} \rho(\rho + 1) (\overline{g}_2 - g_2)\big) (\overline{c}_2 - c_2) +  \frac{1}{6} \rho (\rho + 1)(\overline{g}_1 - g_1)(\overline{c}_3 - c_3)\nonumber \\
&+ \frac{1}{24}\rho^4 (\overline{g}_1 - g_1)^4 + \frac{1}{4} \rho^3c (\overline{g}_1 - g_1)^3 + \Big( - \frac{1}{4}\rho^3 (\overline{g}_2 - g_2)\nonumber \\
&+ \frac{1}{3} \rho (\rho +1) (\frac{11}{8} c^2 - c_2) (\overline{g}_1 - g_1)^2\Big) - \Big( - \frac{7}{12} c \rho (\rho + 1) (\overline{g}_2 - g_2)\nonumber \\
&- \frac{1}{3} \rho ( \rho + 1)  (\overline{g}_3 - g_3)\Big)
(\overline{g}_1 - g_1) + \frac{1}{8} \rho (\rho + 1) (\overline{g}_2 - g_2)^2 \Big]a^4
+ \ldots \bigg\}.
\end{align}
\end{subequations}
Eqs.  (20a,b) satisfy the consistency conditions $\overline{\overline{a}} (\overline{a}(a)) = \overline{\overline{a}} (a)$ and $\overline{\overline{m}} (\overline{m}(m,a), \overline{a} (a)) = \overline{\overline{m}} (m,a)$. 
If $c_i = g_i = 0$, then $a$ and $m$ are in the 't Hooft renormalization scheme and evolve according to $\sigma_n$ and $\tau_n$ in eqs. (12,13).  Using eq. (17) we can determine the value of the running coupling and mass in another scheme (such as $\overline{MS}$) which employ $\overline{c}_i$ and $\overline{g}_i$.  This approach provides an alternative to what we obtained in eq. (10).

The approach we have outlined in this paper has been applied to one-coupling, one-mass model. It is straightforward to extend the approach to models which involve more than one coupling or more than one mass. If, for example, there are $N$ couplings $g_a(a = 1,2 \ldots N)$, these running couplings would satisfy equations of the form 
\begin{align}
\mu \frac{d}{d\mu} g_a &= \beta_a \left( g_1, \ldots, g_N \right) \qquad \qquad (a - 1, \ldots N)\nonumber \\
&= \sum_{p = 2}^\infty \sum_{i_1 = 0}^p \ldots \sum_{i_{N=0}}^p \delta_{p,i_1+i_2 + \ldots + i_N} x_{i_1\ldots i_N}^a 
\left( g_1 \right)^{i_{1}}  \ldots  \left(g_N\right)^{i_{p}} 
\end{align}
\begin{equation}
=\sum_{p = 2}^\infty \sum_{i_1 = 0}^p \sum_{i_2 = 0}^{p-i_1}\ldots 
\sum_{i_{N-1}=0}^{p-i_1\ldots -i_{N-2}} x_{i_1,i_2\ldots i_{N-1}, p-i_1-i_2\ldots -i_{N-1}}^a (g_1)^{i_1} (g_2)^{i_2} \ldots (g_N)^{p-i_1-i_2\ldots -i_{N-1}}\nonumber
\end{equation}
when using mass independent renormalization.  This is a generalization of eq. (1).  Only in exceptional circumstances can this equations be integrated analytically, even when working to just one-loop order [10].

However, in analogy with eq. (4a), we can express
\begin{equation}
g^\prime_a = g_a  +  \sum_{n=2}^\infty\sum_{p=1}^{n-1} \sum_{i_{1}=0}^\infty \ldots \sum_{i_{N} = 0}^\infty \delta_{n, i_1 + i_2 \ldots + i_n} 
\alpha_{p; n; i_1 \ldots i_N}^a \ell^p g_1^{i_1} \ldots g_N^{i_N}
\end{equation}
where again $\ell = \ln \left(\frac{\mu}{\mu^\prime}\right)$, $g_a^\prime = g_a(\mu^\prime)$ and $g_a = g_a(\mu)$.  The equations
\begin{equation}
\mu \frac{d}{d\mu} g^\prime_a = \left( \mu \frac{\partial}{\partial\mu} + \beta_b \frac{\partial}{\partial g_n}\right) g^\prime_a = 0
\end{equation}
can be used to find $\alpha_{p;n; i_1 \ldots i_N}^a$ in terms of  $x_{p; i_1 \ldots i_N}^a$.  It is immediately apparent that $x_{2j}^a = -\alpha_{1,2,2j}^a$ if there are two couplings. Similar considerations can be used if there are multiple masses $m_a(a = 1 \ldots M)$ to show how they evolve when the renormalization mass scale $\mu$ changes.  A discussion of how $g_a$ varies under a change of RS appears in refs. [13,16]. 

In Fig. 1 we plot the result of eq. (6a), cutting the infinite sum off at $n=0, 1, 2, 3$ and $4$. This graph has been obtained using the boundary value $a(M_z)=0.1185/\pi$ and $n_f=5$ flavours. These curves imply that including more terms in the sum of eq. (6a) leads to successively closer approximations to the exact behaviour of the coupling.

The three curves in Fig. 2 present results from three different ways of obtaining $a(\mu)$ from eq. (1a) after $b, c, c_2, c_3$ and $c_4$ are computed. For the curve obtained by using the sum of eq. (6a) up to $n=4$, we use the boundary value of $a(m_b = 4.18 GeV)=0.072121836$. The same boundary condition is used on the second curve which is found by numerical integration of eq. (6a) using the flve-loop approximation to $\beta(a)$ when using $\overline{MS}$. A third curve is obtained using the five loop result which follows from eq. (B8) using a value of $\Lambda_{QCD}$ in $L$ which is consistent with $n_f=5$ flavours and $a(m_b = 4.18 GeV)=0.072121836$.. By construction these curves for a($\mu$) all intersect only when $\mu=m_b$, though all three decrease as $\mu$ increases and are concave upwards.  

Curves for the running mass $m(\mu)$ are presented in Fig. 3. We see that if $m_b(m_b) = 4.18 GeV$, then all of the curves obtained by perfoming the sum in eq. (6b) with cutoffs at $n = 0, 1, 2, 3$ and $4$ coincide while direct integration of the five loop approximationo of eq. (1b) is distinct (though all of these graphs are decreasing at a decreasing rate as $\mu$ increases). 

When examining a perturbative evaluation of a physical process using the functional dependence of the running coupling and running mass on the mass scale, it is apparent from our figures that using the RG summed results developed in our paper are to be preferred.  In Fig. 1 we see that RG summation, by virtue of the fact that it includes contributions from all orders of perturbation theory, is relatively insensitive to inclusion of higher loop effects. We also see that ignoring the effects of higher orders, even when working with the five loop beta function, leads to distinct values of the running coupling. Fig. 2 shows that RG summation leads to a distinct value for the running coupling at high mass scale; it differs not only from what comes from direct integration of the defining equation for the running coupling, but also from the result of using the usual approach to obtaining the running coupling outlined in appendix B. This improvement is clearly the consequence of incorporating the contribution of higher loop effects through use of RG summation. Finally, Fig. 3 shows how RG summation of higher loop effects into the dependence of the running mass on the renormalization mass scale leads to results that are distinct from simply using the five loop result for the anomalous mass dimension and that these RG summed results are not greatly affected by higher loop contributions. To further substantiate these present findings, we are investigating distinct physical processes in an upcoming work [18]. 
\\

{\Large\bf{Acknowledgements}}\\
R. Macleod had a helpful comment.

\setcounter{equation}{0}

\section*{Appendix A}

The full solutions for $T_3$ and $T_4$ obtained from eq. (7b) are
\begin{align}
T_3 =\frac{1}{6}\rho\,{w}^{\rho-3}[C_{{3,0}} \ln ^{3} w + \left( C_{{2,1}}w+C_{{2,0}} \right) \ln^{2}w + \left( C_{{1,2}}{w}^{2}+C_{{1,1}}w+C_{{1,0}} \right) \ln w \nonumber \\
+ C_{{0,3}}{w}^{3}+C_{{0,2}}{w}^{2}+C_{{0,1}}w+C_{{0,0}}] \tag{A.1}
\end{align}

\begin{align}
T_{{4}}=\frac{1}{24}\rho\,{w}^{\rho-4} [ D_{{4,0}} \ln ^{4} w + \left( D_{{3,1}}w+D_{{3,0}} \right)  \ln ^{3}w + ( D_{{2,2}}{w}^{2}+D_{{2,1}}w 
+D_{{2,0}} )  \ln ^{2} w  \nonumber \\
+( D_{{1,3}}{w}^{3}+D_{{1,2}}{w}^{2}+D_{{1,1}}w+D_{{1,0}}) \ln w +D_{{0,4}}{w}^{4}+D_{{0,3}}{w}^{3}+D_{{0,2}}{w}^{2}+D_{{0,1}}w+D_{{0,0}}] \tag{A.2}
\end{align}

where the associated coefficients for $T_3$ solution are 
\begin{equation}\tag{A.3}
C_{{0,0}}= \left( c-g_{{1}} \right) ^{3}{\rho}^{2}+ \left( 3\,c-3\,g_{{1}} \right)  \left( {c}^{2}+cg_{{1}}-c_{{2}}-g_{{2}} \right) \rho-{c}^{3}+4\,{c}^{2}g_{{1}}+ \left( 2\,c_{{2}}+2\,g_{{2}} \right) c-4\,c_{{2}}g_{{1}}-c_{{3}}-2\,g_{{3}} 
\end{equation}

\begin{equation}\tag{A.4}
C_{{0,1}}=-3\, \left( c-g_{{1}} \right) ^{3}{\rho}^{2}- \left( 3\,c-3\,g_{{1}} \right)  \left( 3\,{c}^{2}+cg_{{1}}-3\,c_{{2}}-g_{{2}}\right) \rho-6\,{c}^{2}g_{{1}}+6\,c_{{2}}g_{{1}} 
\end{equation}

\begin{equation}\tag{A.5}
C_{{0,2}}=3\, \left( c-g_{{1}} \right) ^{3}{\rho}^{2}+ \left( 3\,c-3\,g_{{1}} \right)  \left( 3\,{c}^{2}-cg_{{1}}-3\,c_{{2}}+g_{{2}}\right) \rho+3\,{c}^{3}-6\,cc_{{2}}+3\,c_{{3}}
\end{equation}

\begin{equation}\tag{A.6}
C_{{0,3}}=- \left( c-g_{{1}} \right) ^{3}{\rho}^{2}- \left( 3\,c-3\,g_{{1}} \right)  \left( {c}^{2}-cg_{{1}}-c_{{2}}+g_{{2}} \right) \rho-2\,{c}^{3}+2\,{c}^{2}g_{{1}}+ \left( 4\,c_{{2}}-2\,g_{{2}} \right) c-2\,c_{{2}}g_{{1}}-2\,c_{{3}}+2\,g_{{3}}
\end{equation}

\begin{equation}\tag{A.7}
C_{{1,0}}=3\,c \left[  \left( c-g_{{1}} \right) ^{2}{\rho}^{2}+\left( {c}^{2}+3\,cg_{{1}}-2\,{g_{{1}}}^{2}-c_{{2}}-g_{{2}} \right) \rho-2\,{c}^{2}+2\,c_{{2}}+2\,g_{{2}} \right]
\end{equation}

\begin{equation}\tag{A.8}
C_{{1,1}}=-6\,c \left[  \left( c-g_{{1}} \right) ^{2}{\rho}^{2}+\left( {c}^{2}+cg_{{1}}-{g_{{1}}}^{2}-c_{{2}} \right) \rho-{c}^{2}+c_{{2}} \right]
\end{equation}

\begin{equation}\tag{A.8}
C_{{1,2}}=3\,c \left[  \left( c-g_{{1}} \right) ^{2}{\rho}^{2}+\left( {c}^{2}-cg_{{1}}-c_{{2}}+g_{{2}} \right) \rho \right]
\end{equation}

\begin{equation}\tag{A.9}
C_{{2,0}}=3\,{c}^{2} \left[  \left( c-g_{{1}} \right) {\rho}^{2}+\left( -c+3\,g_{{1}} \right) \rho-c-2\,g_{{1}} \right] 
\end{equation}

\begin{equation}\tag{A.10}
C_{{2,1}}=-3\,\rho\,{c}^{2} \left( \rho-1 \right)  \left( c-g_{{1}} \right)
\end{equation}

\begin{equation}\tag{A.11}
C_{{3,0}}={c}^{3} \left( \rho-1 \right)  \left( \rho-2 \right) 
\end{equation}.

The coefficients for the $T_4$ solution are 

\begin{align}
D_{{0,0}}={\rho}^{3} ( c-g_{{1}} ) ^{4}+6\,{\rho}^{2} ( c-g_{{1}} ) ^{2} ( {c}^{2}+cg_{{1}}-c_{{2}}-g_{{2}} ) +\rho\, ( -{c}^{4}+26\,{c}^{3}g_{{1}}+ ( -13\,{g_{{1}}}^{2}+2\,c_{{2}}+2\,g_{{2}} ) {c}^{2} \nonumber \\
+( -30\,c_{{2}}g_{{1}}-14\,g_{{1}}g_{{2}}-4\,c_{{3}}-8\,g_{{3}}) c+3\,{c_{{2}}}^{2}+ ( 16\,{g_{{1}}}^{2}+6\,g_{{2}} ) c_{{2}}+4\,c_{{3}}g_{{1}}+8\,g_{{1}}g_{{3}}+3\,{g_{{2}}}^{2} ) -10\,{c}^{4}-6\,{c}^{3}g_{{1}} \nonumber \\
+ ( 18\,c_{{2}}+18\,g_{{2}}) {c}^{2}+ ( 12\,c_{{2}}g_{{1}}+4\,c_{{3}}+6\,g_{{3}} ) c-10\,{c_{{2}}}^{2}-18\,c_{{2}}g_{{2}}-6\,c_{{3}}g_{{1}}-2\,c_{{4}}-6\,g_{{4}}\tag{A.12}
\end{align}

\begin{align}
D_{{0,1}}=-4\,{\rho}^{3} ( c-g_{{1}} ) ^{4}-12\,{\rho}^{2}( c-g_{{1}}) ^{2} ( 2\,{c}^{2}+cg_{{1}}-2\,c_{{2}}-g_{{2}} ) +\rho\, ( -8\,{c}^{4}-56\,{c}^{3}g_{{1}}+ ( 40\,{g_{{1}}}^{2}+16\,c_{{2}}+ 4\,g_{{2}} ) {c}^{2}\nonumber \\
+ ( 60\,c_{{2}}g_{{1}}+8\,g_{{1}}g_{{2}} + 4\,c_{{3}}+8\,g_{{3}} ) c-12\,{c_{{2}}}^{2} + ( -40\,{g_{{1}}}^{2}-12\,g_{{2}} ) c_{{2}} - 4\,c_{{3}}g_{{1}} - 8\,g_{{1}}g_{{3}} ) \nonumber \\
+ ( 24\,{c}^{2}-24\,c_{{2}} ) ( {c}^{2}-c_{{2}}-g_{{2}} )\tag{A.13}
\end{align}

\begin{align}
D_{{0,2}}=6\,{\rho}^{3} ( c-g_{{1}}) ^{4}+36\,{\rho}^{2}
 \left( c-g_{{1}} \right) ^{2} \left( {c}^{2}-c_{{2}} \right) +\rho\,
( 30\,{c}^{4}+12\,{c}^{3}g_{{1}}+ \left( -30\,{g_{{1}}}^{2}-60\,
c_{{2}} \right) {c}^{2}+ \left( 12\,g_{{1}}g_{{2}}+12\,c_{{3}}\right) c \nonumber \\
+24\,c_{{2}}{g_{{1}}}^{2}+18\,{c_{{2}}}^{2}-12\,c_{{3}}g_{{1
}}-6\,{g_{{2}}}^{2} ) -12\,{c}^{4}+12\,{c}^{3}g_{{1}}+24\,{c}^{2
}c_{{2}}-24\,cc_{{2}}g_{{1}}-12\,{c_{{2}}}^{2}+12\,c_{{3}}g_{{1}}\tag{A.14}
\end{align}

\begin{align}
D_{{0,3}}=-4\,{\rho}^{3} \left( c-g_{{1}} \right) ^{4}-12\,{\rho}^{2}
 \left( c-g_{{1}} \right) ^{2} \left( 2\,{c}^{2}-cg_{{1}}-2\,c_{{2}}+g
_{{2}} \right) +\rho\,( -32\,{c}^{4}+40\,{c}^{3}g_{{1}}+ ( 
-8\,{g_{{1}}}^{2} +64\,c_{{2}}\nonumber \\
-20\,g_{{2}}){c}^{2} + ( -60\,c_{{2}}g_{{1}}+8\,g_{{1}}g_{{2}}-20\,c_{{3}}
+8\,g_{{3}} ) c-12\,{c_{{2}}}^{2}+ \left( 8\,{g_{{1}}}^{2}+12\,g_{{2}} \right) c_{{2}}+
20\,c_{{3}}g_{{1}}-8\,g_{{1}}g_{{3}}) \nonumber \\
-8\,{c}^{4}+24\,{c}^{2}c_{{2}}-16\,cc_{{3}}-8\,{c_{{2}}}^{2}+8\,c_{{4}}\tag{A.15}
\end{align}

\begin{align}
D_{{0,4}}={\rho}^{3} \left( c-g_{{1}} \right) ^{4}+6\,{\rho}^{2}
 \left( c-g_{{1}} \right) ^{2} \left( {c}^{2}-cg_{{1}}-c_{{2}}+g_{{2}}
 \right) +\rho\,( 11\,{c}^{4}-22\,{c}^{3}g_{{1}}+ \left( 11\,{g_
{{1}}}^{2}-22\,c_{{2}}+14\,g_{{2}} \right) {c}^{2}\nonumber \\
+ \left( 30\,c_{{2}}
g_{{1}}-14\,g_{{1}}g_{{2}}+8\,c_{{3}}-8\,g_{{3}} \right) c-8\,c_{{2}}{
g_{{1}}}^{2}+3\,{c_{{2}}}^{2}-6\,c_{{2}}g_{{2}}-8\,c_{{3}}g_{{1}}+8\,g
_{{1}}g_{{3}}+3\,{g_{{2}}}^{2}) +6\,{c}^{4}-6\,{c}^{3}g_{{1}} \nonumber \\
+ \left( -18\,c_{{2}}+6\,g_{{2}} \right) {c}^{2}+ \left( 12\,c_{{2}}g_{
{1}}+12\,c_{{3}}-6\,g_{{3}} \right) c+6\,{c_{{2}}}^{2}-6\,c_{{2}}g_{{2
}}-6\,c_{{3}}g_{{1}}-6\,c_{{4}}+6\,g_{{4}}\tag{A.16}
\end{align}

\begin{align}
D_{{1,0}}=4\,c [  \left( c-g_{{1}} \right) ^{3}{\rho}^{3}+
 \left( 3\,c-3\,g_{{1}} \right)  \left( {c}^{2}+2\,cg_{{1}}-{g_{{1}}}^
{2}-c_{{2}}-g_{{2}} \right) {\rho}^{2}+ ( -7\,{c}^{3}+13\,{c}^{2}
g_{{1}}\nonumber \\
+ \left( 3\,{g_{{1}}}^{2}+8\,c_{{2}}+8\,g_{{2}} \right) c-13\,c
_{{2}}g_{{1}}-9\,g_{{1}}g_{{2}}-c_{{3}}-2\,g_{{3}}) \rho-3\,{c}
^{3}-12\,{c}^{2}g_{{1}}+12\,c_{{2}}g_{{1}}+3\,c_{{3}}+6\,g_{{3}}] \tag{A.17}
\end{align}

\begin{align}
D_{{1,1}}=-12\,c [  \left( c-g_{{1}} \right) ^{3}{\rho}^{3}+
 \left( c-g_{{1}} \right)  \left( 3\,{c}^{2}+3\,cg_{{1}}-2\,{g_{{1}}}^
{2}-3\,c_{{2}}-g_{{2}} \right) {\rho}^{2}+( -4\,{c}^{3}+8\,{c}^{2}g_{{1}}
+4\,cc_{{2}}+2\,cg_{{2}}\nonumber \\
-8\,c_{{2}}g_{{1}}-2\,g_{{1}}g_{{2}}) \rho-2\,{c}^{3}-4\,{c}^{2}g_{{1}}+2\,cc_{{2}}+4\,c_{{2}}g_{{1
}}] \tag{A.18}
\end{align}

\begin{align}
D_{{1,2}}=12\,c [\left( c-g_{{1}} \right) ^{3}{\rho}^{3}+
 \left( c-g_{{1}} \right)  \left( 3\,{c}^{2}-{g_{{1}}}^{2}-3\,c_{{2}}+
g_{{2}} \right) {\rho}^{2}+ ( -{c}^{3}+3\,{c}^{2}g_{{1}}-c{g_{{1}
}}^{2}-3\,c_{{2}}g_{{1}}\nonumber \\
+g_{{1}}g_{{2}}+c_{{3}}) \rho-{c}^{3}+2\,cc_{{2}}-c_{{3}}] \tag{A.19}
\end{align}

\begin{align}
D_{{1,3}}=-4\,c\rho\, [  \left( c-g_{{1}} \right) ^{3}{\rho}^{2}+\left( 3\,c-3\,g_{{1}} \right)  \left( {c}^{2}-cg_{{1}}-c_{{2}}+g_{{2}} \right) \rho+2\,{c}^{3}-2\,{c}^{2}g_{{1}}+ \left( -4\,c_{{2}}+2\,g_
{{2}} \right) c \nonumber \\
+2\,c_{{2}}g_{{1}}+2\,c_{{3}}-2\,g_{{3}} ] \tag{A.20}
\end{align}

\begin{align}
D_{{2,0}}=6\,{c}^{2}  [ \left( c-g_{{1}} \right) ^{2}{\rho}^{3}+ \left( 7\,cg_{{1}}-5\,{g_{{1}}}^{2}-c_{{2}}-g_{{2}} \right) {\rho}^{2} + ( -7\,{c}^{2}-3\,cg_{{1}}+6\,{g_{{1}}}^{2}+5\,c_{{2}}+5\,g_{{2}} ) \rho \nonumber \\
+4\,{c}^{2}-4\,cg_{{1}}-6\,c_{{2}}-6\,g_{{2}} ] \tag{A.21}
\end{align}

\begin{align}
D_{{2,1}}=-12\,{c}^{2} \left(  \left( c-g_{{1}} \right) ^{2}{\rho}^{3}
+ \left( 4\,cg_{{1}}-3\,{g_{{1}}}^{2}-c_{{2}} \right) {\rho}^{2}+
 \left( -4\,{c}^{2}-cg_{{1}}+2\,{g_{{1}}}^{2}+3\,c_{{2}} \right) \rho+
2\,{c}^{2}-2\,c_{{2}} \right) \tag{A.22}
\end{align}

\begin{align}
D_{{2,2}}=6\,{c}^{2}\rho\, \left( \rho-1 \right)  \left(  \left( c-g_{
{1}} \right) ^{2}\rho+c \left( c-g_{{1}} \right) -c_{{2}}+g_{{2}}
 \right) \tag{A.23}
\end{align}

\begin{align}
D_{{3,0}}=4\,{c}^{3} \left(  \left( c-g_{{1}} \right) {\rho}^{3}+
 \left( -3\,c+6\,g_{{1}} \right) {\rho}^{2}+ \left( -c-11\,g_{{1}}
 \right) \rho+5\,c+6\,g_{{1}} \right) \tag{A.24}
\end{align}

\begin{align}
D_{{3,1}}=-4\,{c}^{3}\rho\, \left( \rho-1 \right)  \left( \rho-2\right)  \left( c-g_{{1}} \right)\tag{A.25}
\end{align}

\begin{align}
D_{{4,0}}={c}^{4} \left( \rho-1 \right)  \left( \rho-2 \right) \tag{A.26}
 \left( \rho-3 \right) 
\end{align}

\section*{Appendix B}
\setcounter{equation}{0}
The equation
\begin{equation}\tag{B.1}
a(\mu^\prime) = \sum_{n=0}^\infty S_n(a\ell)a^{n+1}
\end{equation}
which follows from eqs. (4a,6a) shows how $a$ evolves from a boundary value $a(\mu)$ to a value $a(\mu^\prime)$. An alternate way of describing the way in which $a$ evolves under a change in mass scale is described in refs. [14,15].  Here systematic approximations are used to integrate eq. (3a).  An alternate approach is to make the ansatz
\begin{equation}\tag{B.2}
a(\mu) = \sum_{n=1}^\infty \sum_{m=0}^{n-1} \frac{x_{nm}\ln^mL}{L^n} \qquad \left(L \equiv \ln \frac{\mu}{\Lambda}\right).
\end{equation}
With $\beta(a)$ given by eq. (1a), we see that, for example, at order $\frac{1}{L^2}$
\begin{equation}\tag{B.3}
-x_{10} = -b x_{10}^2 \Rightarrow x_{10} = 1/b.
\end{equation}
We find from higher order terms that
\begin{equation}\tag{B.4}
x_{21} = \frac{-c}{b^2} - x_{20}
\end{equation}
\begin{equation}\tag{B.5}
x_{31} - x_{30} = -bx_{20}^2 + \frac{3x_{20}}{b} - \frac{3c}{b} x_{20} - \frac{c_2}{b^3}
\end{equation}
\begin{equation}\tag{B.6}
2x_{32} - 3x_{31} = \frac{5cx_{20}}{b} + 2b x_{20}^2 - 2x_{31} + \frac{3c^2}{b^3}
\end{equation}
\begin{equation}\tag{B.7}
3x_{32} = \frac{c^2}{b^3} + \frac{2x_{20}c}{b} + bx_{20}^2 + 2 x_{32}
\end{equation}
etc.

The boundary condition chosen in refs. [14,15] amounts to setting $x_{20} = 0$; eq. (B.2) leads to 
\begin{align}\tag{B.8}
a(\mu)  = \frac{1}{bL} &- \frac{c\ln L}{b^2L^2} + \frac{1}{b^3L^3}\left[ c^2 (\ln^2L - \ln L - 1) + c_2^2\right]\\
&+ \frac{1}{b^4L^4}\left[ c^3 \left(-\ln^3L + \frac{5}{2} \ln^2L + 2\ln L - \frac{1}{2}\right) - 3 cc_2\ln L + \frac{1}{2}c_3\right]\nonumber \\
& + \frac{1}{b^5L^5}\Bigg[ c^4 \left( \ln^4L - \frac{13}{3}\ln^3 L - \frac{3}{2} \ln^2 L + 4 \ln L + \frac{7}{6}\right)\nonumber \\
& + c^2c_2 (6\ln^2L - 3\ln L -3) - cc_3 \left(2\ln L + \frac{1}{6}\right) + \frac{1}{3} (5c_2^2 + c_4) \Bigg] + \ldots .\nonumber
\end{align}
This is quite distinct from eq. (B.1).

\begin{figure}[hbt]
\begin{center}
\includegraphics[scale=0.95]{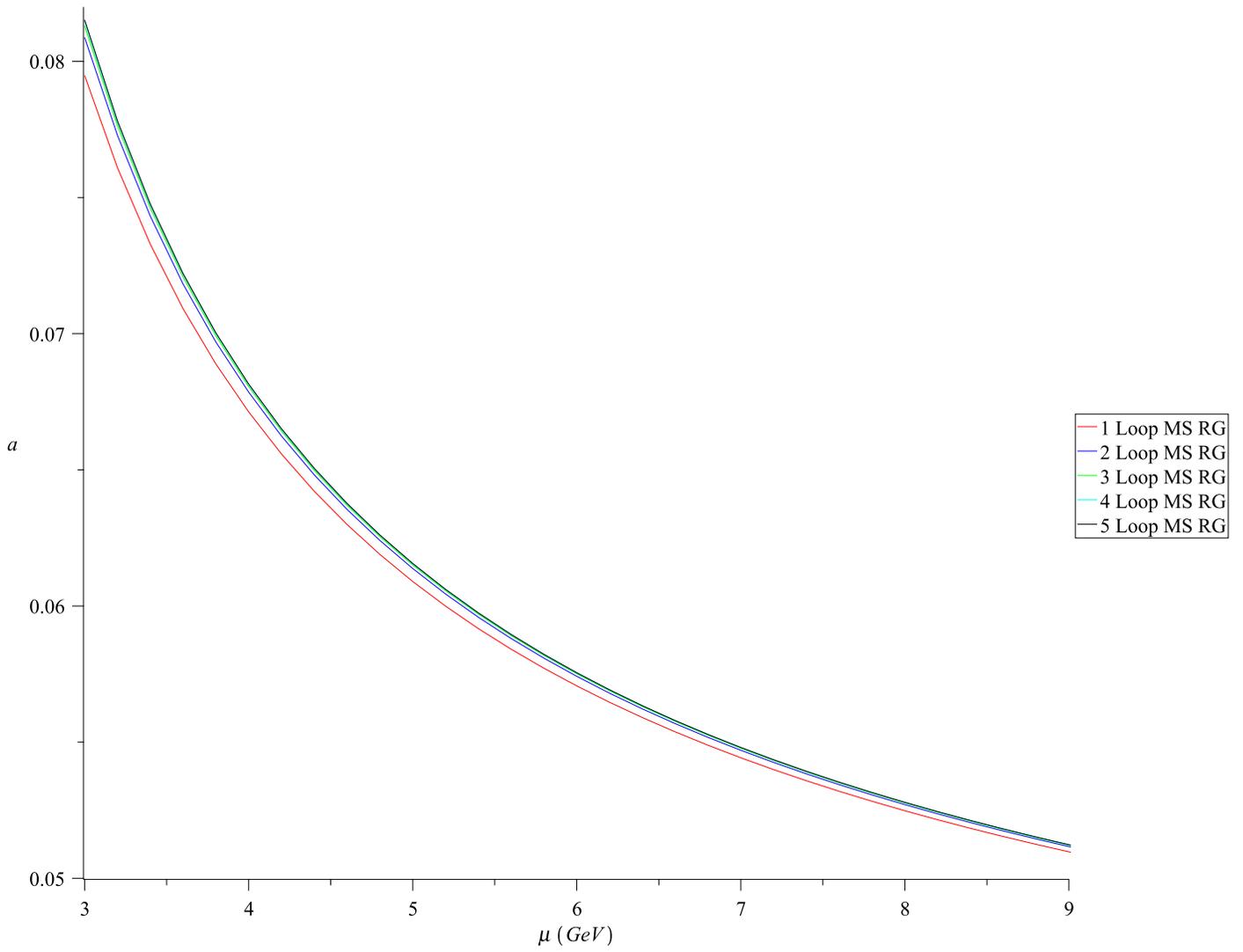}
\caption{The $\mu$ dependence of coupling $a$ from 1 to 5-loop RG summation expressions in the ${\overline{MS}}$ scheme}
\label{Fig. 1}
\end{center}
\end{figure}

\begin{figure}[hbt]
\begin{center}
\includegraphics[scale=0.95]{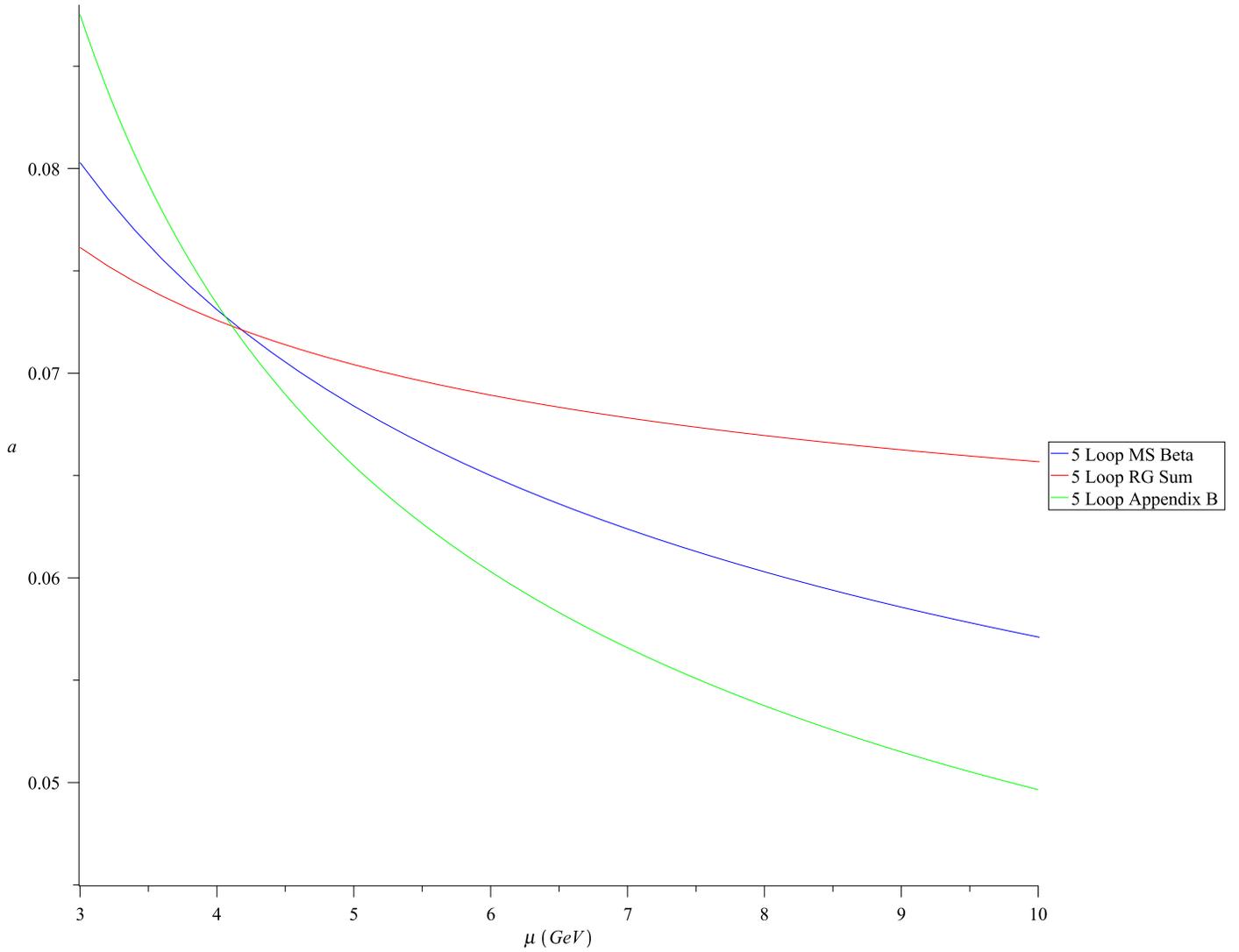}
\caption{The $\mu$ dependence of running coupling $a$ as determined from 5-loop RG $\beta$ function in comparison with 5-loop RG summation expression and 5-loop expansion in Appendix B in the ${\overline{MS}}$ scheme}
\label{Fig. 2}
\end{center}
\end{figure}

\begin{figure}[hbt]
\begin{center}
\includegraphics[scale=0.95]{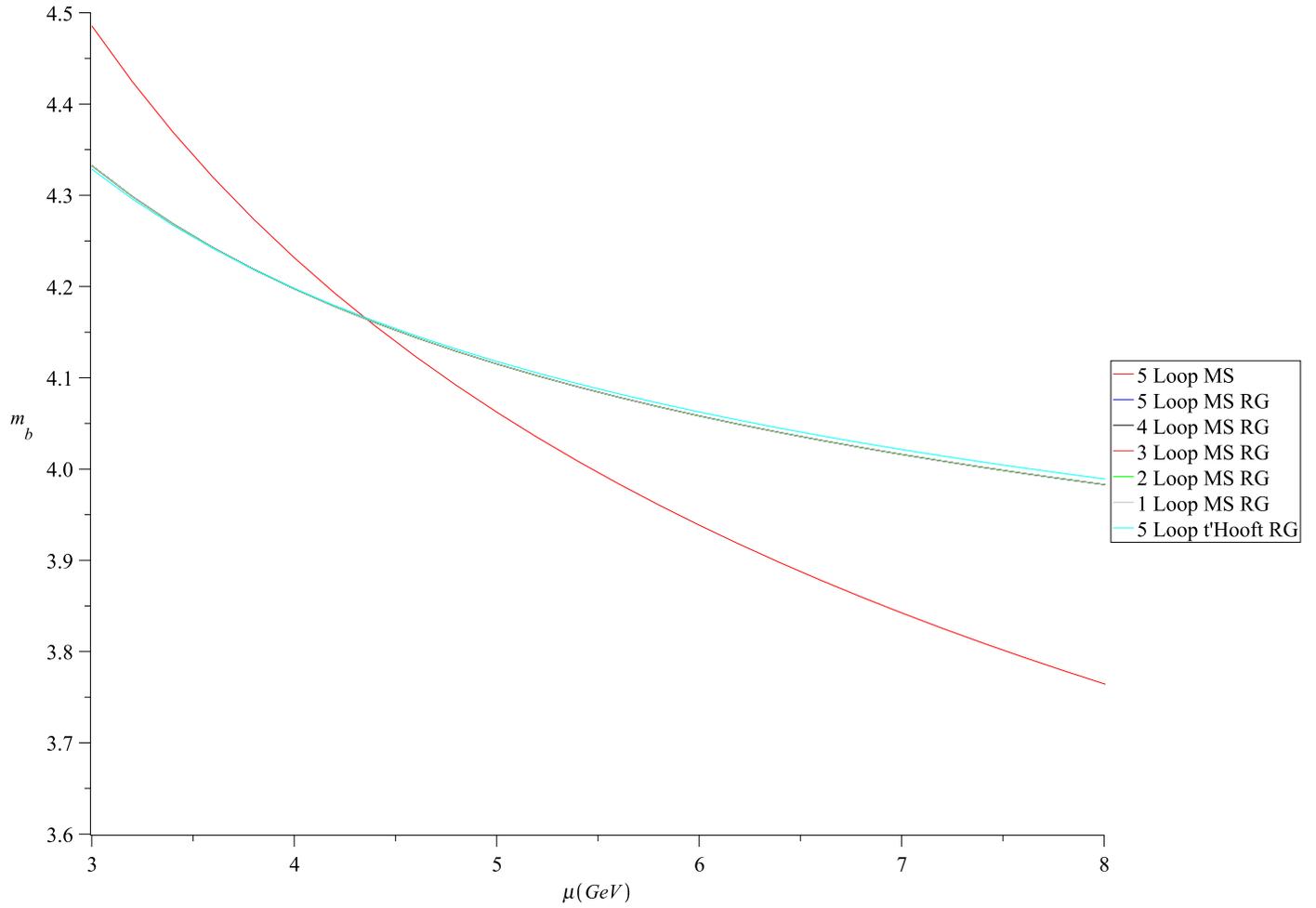}
\caption{The $\mu$ dependence of running b-quark mass from ${\overline{MS}}$ 5-loop RG function as compared to RG summation expressions from 1 to 5 loops in the ${\overline{MS}}$ and t'Hooft schemes}
\label{Fig. 2}
\end{center}
\end{figure}

\end{document}